\documentclass[slac_one]{revtex4}
\usepackage{graphicx}
\usepackage{fancyhdr}
\begin{document}

\title{Dijet Angular Distributions at the ATLAS Experiment} 

\author{P.O. DeViveiros}
\affiliation{University of Toronto}
\author{\emph{on behalf of the ATLAS Collaboration}}

\begin{abstract}
Dijet angular distributions from the first proton-proton collisions at 7 TeV have been
measured with the ATLAS detector at the LHC.  The dataset used corresponds to a total
integrated luminosity of 61 $\pm$ 7 nb$^{-1}$. Comparing the distributions with Monte
Carlo based QCD predictions shows good agreement between data and QCD.  The distributions
are used to set limits on the existence of quark compositeness.

\end{abstract}

\maketitle

\thispagestyle{fancy}

\section{The Observables}

The existence of quark compositeness would manifest itself as a deviation from Standard Model
QCD predictions at a regime approaching the compositeness scale $\Lambda$.  Dijet observables,
such as, for example, the dijet invariant mass, can be probed to search for such deviations.
The understanding of the jet energy scale is a large source of uncertainty for early data analyses
using dijet final states.  The dependence of the measurement on the jet energy scale can however be
greatly lowered by probing the angular distributions of dijet events.  These distributions still
show clear deviations from QCD predictions in the presence of quark compositeness, due to the
different angular properties of both processes. In this analysis, we use two different observables
to probe the angular distributions, $\chi$ and $R_{C}$.

The observable $\chi$ is defined as:
\begin{equation}
\chi = e^{\left| y_1 - y_2 \right|}
\end{equation}
where $y_1$ and $y_2$ are the rapidities of the two leading jets.  For the dominating t-channel
gluon-exchange QCD processes, the differential cross-section $\frac{d\sigma}{d\chi}$ changes slowly
as a function of $\chi$.  On the other hand, new physics is expected to be more isotropic, leading
to large excesses at low values of $\chi$ for invariant dijet masses approaching the scale of compositeness,
$\Lambda$.  Therefore, this variable offers sensitivity to such new physics.

The dijet centrality ratio $R_{C}$ is defined as the ratio of the number of events where both jets are central
to the number of events where both jets are less central:
\begin{equation}
R_{C} = \frac{N(\left|\eta_{j_1j_2} \right| < 0.5)}{N(0.5 < \left|\eta_{j_1j_2} \right| < 1.0)}
\end{equation}
where $\eta_{j_1j_2}$ is the pseudo-rapidity of both leading jets.  The $R_{C}$ observable is plotted as a
function of the dijet invariant mass, and is expected to be flat for QCD. The presence of quark compositeness
would increase the production of central jets, and would therefore lead to a rise in $R_{C}$ at masses
approaching $\Lambda$.  Because the inner and outer regions can be defined freely, this method can be used
to focus on the detector regions that are best understood, making it ideal for an early measurement.

\begin{figure*}
\centering
\includegraphics[width=65mm]{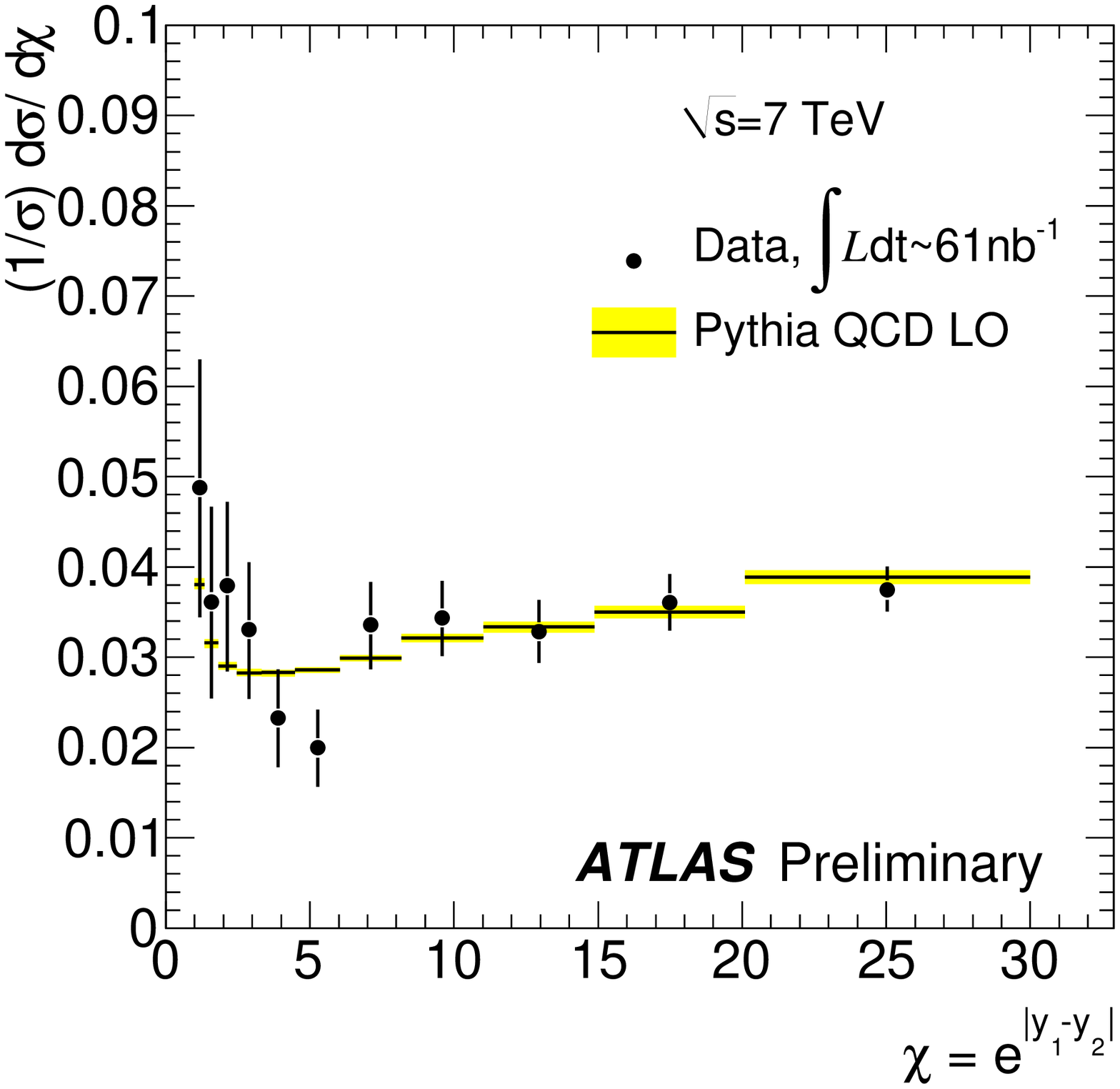}
\includegraphics[width=65mm]{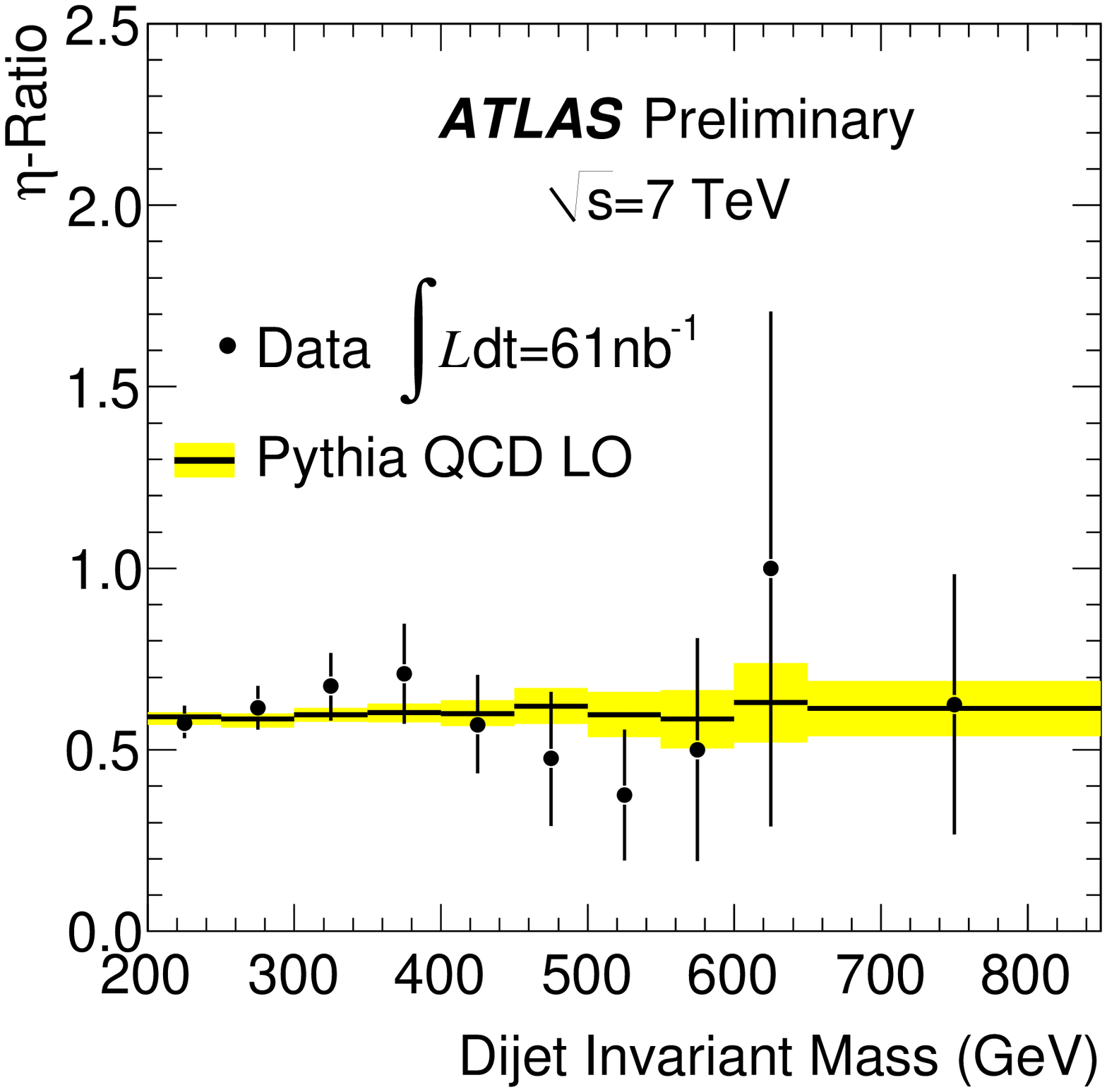}
\caption{Comparison between data and Monte Carlo for the $\chi$ distribution in the 540 $< m_{jj} <$ 680 range (left) and for the $R_{C}$ observable (right).} \label{data}
\end{figure*}

\section{Reconstruction and Event Selection}
Jets are reconstructed in the ATLAS detector \cite{atlas} using the anti-$k_{T}$  algorithm \cite{antikt}, with a size parameter of 0.6.
The inputs to the algorithm are uncalibrated (EM-scale) noise-suppressed 3D energy clusters (topoclusters).
The jets are later calibrated to the hadronic scale to counteract the non-compensating nature of the ATLAS 
calorimeters.  The calibration scheme uses a simple function to restore the jet energy response based 
on the jet p$_{T}$  and $\eta$ \cite{jets}.

We require the events to have at least 2 jets.  The leading jet must have p$_{T}$ larger than 60 GeV and
the subleading jet p$_{T}$ larger than 30 GeV.  These criteria ensure that the jet reconstruction and
used triggers are close to 100\% efficient \cite{jets}.  Furthermore, we reject events where one of the two leading jets is not
associated with an in-time real calorimeter energy deposit.  Also, any event with a jet with p$_{T}$
larger than 15 GeV which could be mismeasured due to detector effects (dead cells, masked cells,
etc.) is rejected. Finally, we require the presence of a well reconstructed event vertex with at least
5 associated tracks.

The $\chi$ observable is studied in two different bins of dijet invariant mass with boundaries (320, 540)
and (540, 680) GeV. The boundaries of the bins were chosen such that the trigger requirements do not bias
the distributions of the invariant mass. Also, the bins in $\chi$ were chosen to have high purity and 
stability: this means that the various detector effects are not expected to cause large bin-to-bin 
migrations for the events.

\begin{figure*}
\centering
\includegraphics[width=65mm]{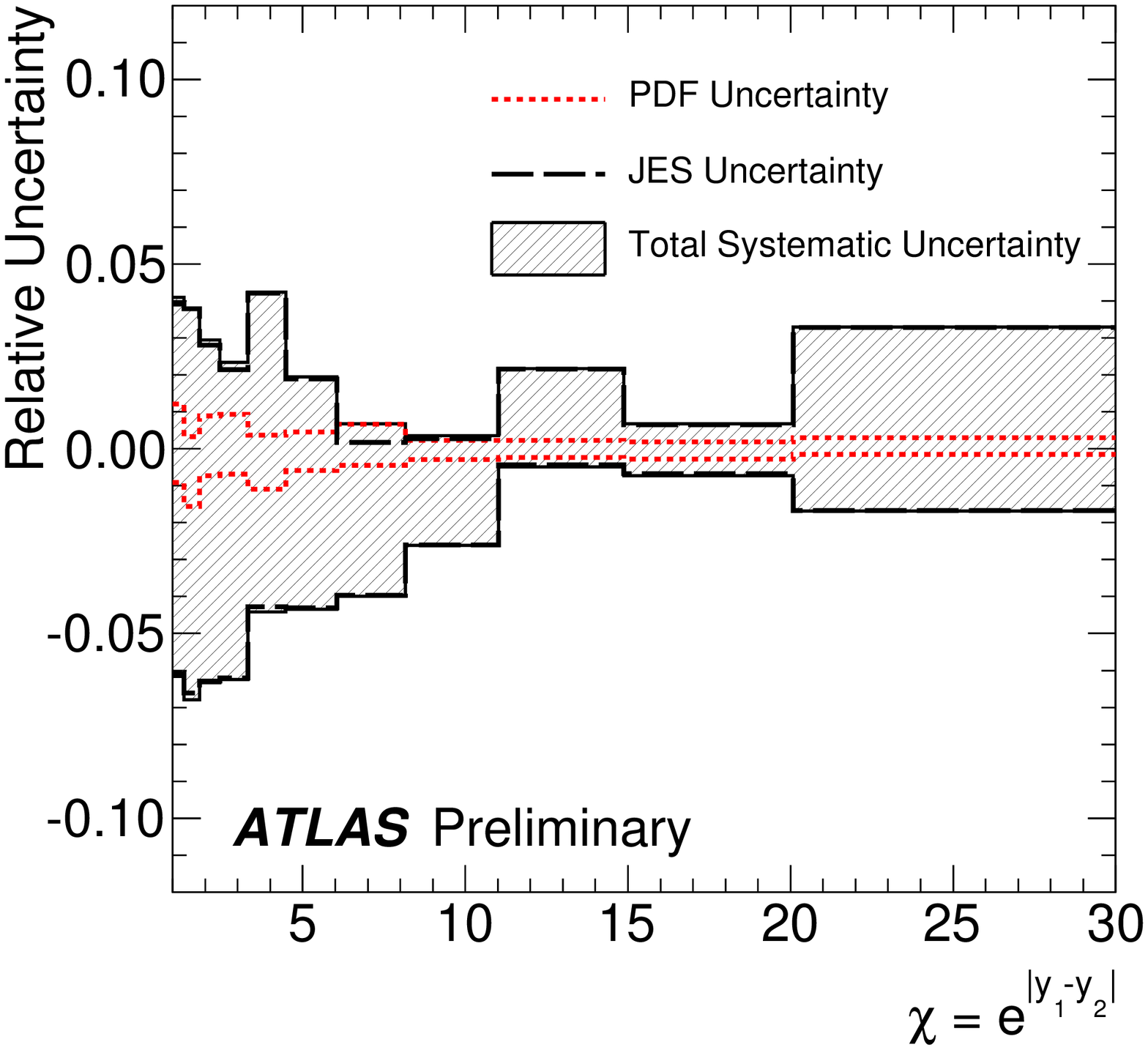}
\includegraphics[width=65mm]{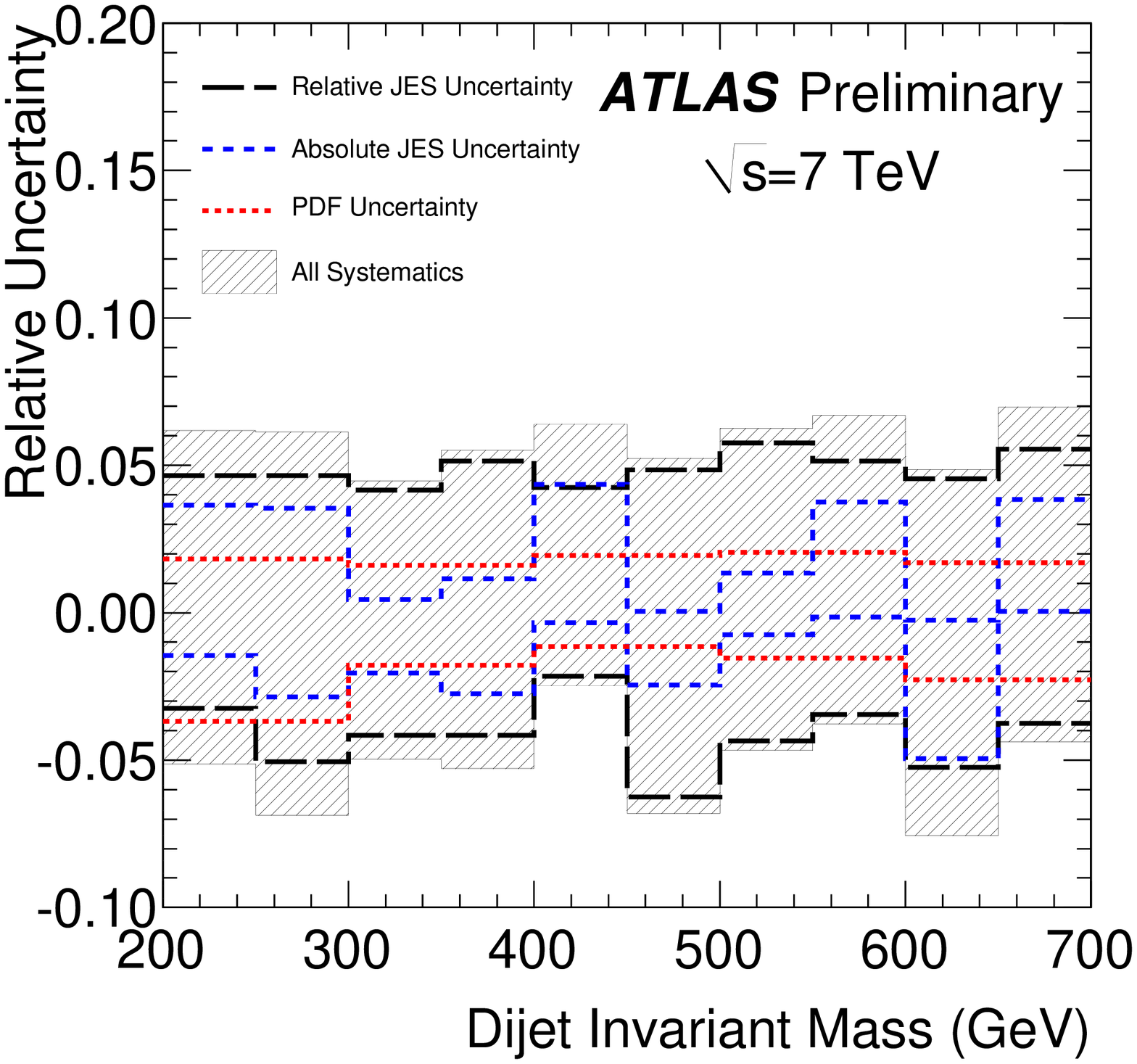}
\caption{Summary of systematic uncertainties for the $\chi$ distribution in the 540 $< m_{jj} <$ 680 range (left) and for the $R_{C}$ observable (right).} \label{syst}
\end{figure*}

\section{The Distributions}
The data is compared to Leading Order Pythia \cite{pythia} QCD Monte Carlo predictions.  A GEANT4
detector full simulation is used to simulate the passage of particles through the detector \cite{geant4}.
The results are shown in Fig. \ref{data}.
Using a simple $\chi^{2}$ test,
the data is found to agree well with the predictions \cite{conf}.

The uncertainties shown on the data are purely statistical.  Various systematic uncertainties have also been
taken into account, and are shown as part of the prediction.  These uncertainties are propagated to the 
observable using a pseudo-experiment based approach.  Shown in Fig. \ref{syst} for both observables are the
systematic uncertanties due to the jet energy scale \cite{jes} and choice of parton distribution functions (PDFs).

\begin{figure*}
\centering
\includegraphics[width=65mm]{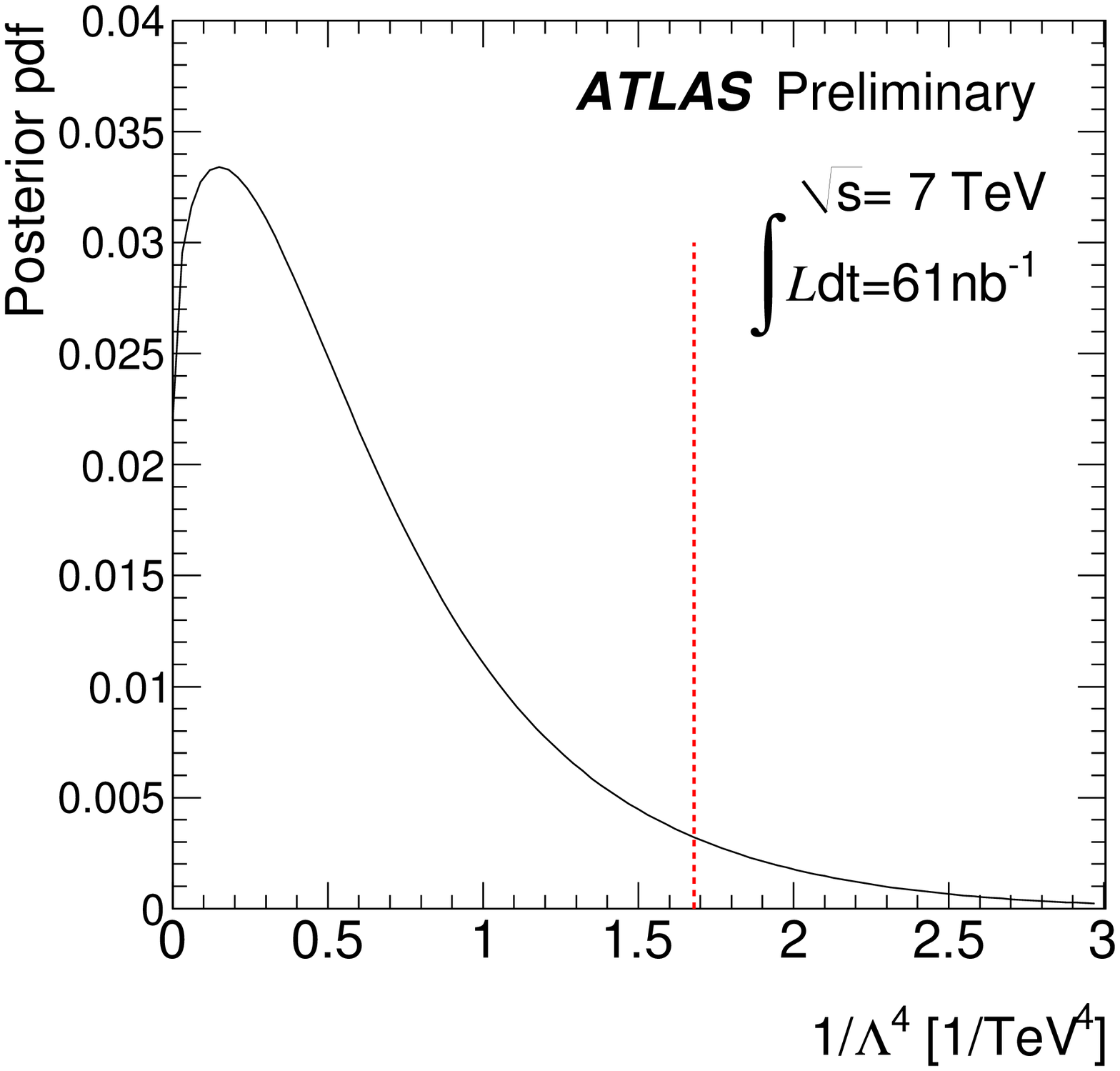}
\includegraphics[width=65mm]{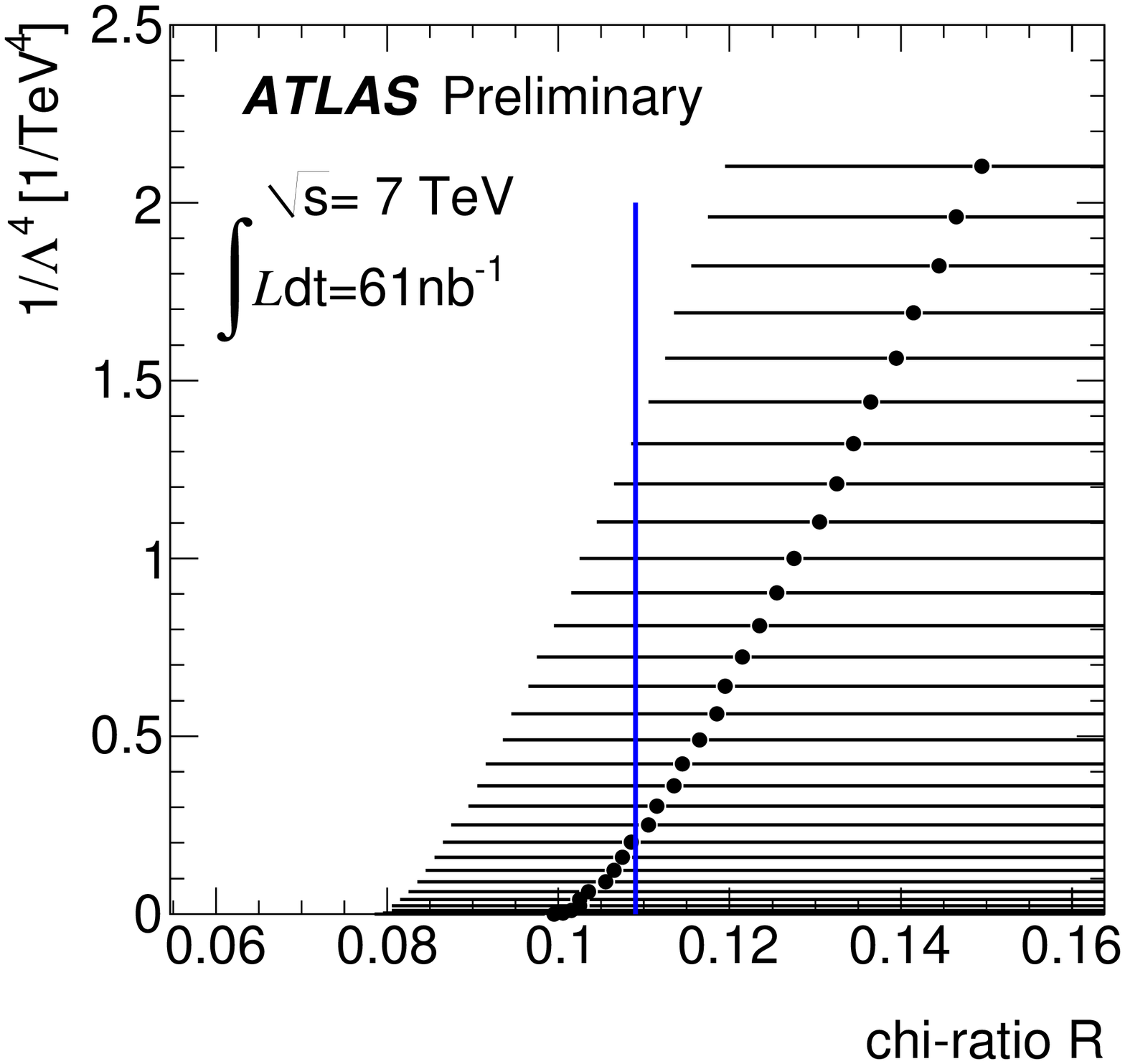}
\caption{Bayesian posterior probability with a flat prior in $\frac{1}{\Lambda^{4}}$ (left) and  Neyman
construction (right) for the $\chi$ observable. The 95\% C.L. limits are shown by the vertical lines.} \label{bayes}
\end{figure*}

\section{Setting Limits}
We set limits on the existence of quark compositeness at a scale $\Lambda$, by using a Fermi four-contact
interaction model.  Systematic uncertainties are taken into account by marginalizing the associated degrees
of freedom.  We set limits using both frequentist and Bayesian approaches.  The probability posterior for the
Bayesian approach, and the Neyman construction for the frequentist approach are both shown in the highest
invariant mass bin for the $\chi$ observable, in Fig. \ref{bayes}.  The associated 95\% C.L. limits on
$\Lambda$ are also shown.  The limits are found to be 875 (930) GeV for the Bayesian (Frequentist) approach
using the $\chi$ observable, and 760 GeV using a Bayesian approach for the $R_{C}$ observable.

\section{Conclusions}
We have measured various dijet angular distributions in the first ATLAS data at 7 TeV using an integrated
luminosity of 61 $\pm$ 7 nb$^{-1}$.  Good agreement is found between the data and the leading order QCD predictions.  
We set limits at the 95\% C.L. on the existence of quark compositeness for a scale lower than 875 GeV using
the $\chi$ spectra, corresponding to a distance scale of 2.3 x 10$^{-4}$ fm.

\end{document}